\documentclass[aps,groupaddress,prx,preprint,showpacs,letterpaper,tighten,float,longbibliography,nofootinbib]{revtex4-1}
\usepackage{amssymb}
\usepackage{amsbsy}
\usepackage{amsmath}
\usepackage{epsfig}
\usepackage{wrapfig}
\usepackage{graphicx}
\usepackage{array}
\usepackage[utf8]{inputenc}
\usepackage{textcomp}
\usepackage{color}
\usepackage{braket}
\usepackage{bm,hyperref}
\usepackage[titletoc,toc,title]{appendix}
\usepackage[normalem]{ulem}

\definecolor{myblue}{rgb}{.93, .93, 1}

\newcommand{\beq}{\begin{equation}}
\newcommand{\eeq}{\end{equation}}

\begin{document}

\title{Fractons}
\author{Rahul M. Nandkishore}
\author{Michael Hermele}

\affiliation{Department of Physics and Center for Theory of Quantum Matter, University of Colorado, Boulder, Colorado 80309, USA}

\date{\today}

\begin{abstract}
We review what is known about fracton phases of quantum matter. Fracton phases are characterized by excitations that exhibit restricted mobility, being either immobile under local Hamiltonian dynamics, or mobile only in certain directions. They constitute a new class of quantum state of matter, which does not wholly fit into any of the existing paradigms, but which connects to areas including glassy quantum dynamics, topological order, spin liquids, elasticity theory, quantum information theory, and gravity. We begin by discussing {\it gapped} fracton phases, which may be described using exactly solvable lattice spin models. We introduce the basic phenomena, and discuss the geometric and topological response of fracton phases. We also discuss connections to generalized gauge theories, and explain how gapped fracton phases may be obtained from more familiar theories. We then introduce the framework of {\it tensor gauge theory}, which provides a powerful complementary perspective on fracton phases. We discuss how tensor gauge theory encodes the fracton phenomenon, and how it allows us to access {\it gapless} fracton phases. We discuss the basic properties of gapless fracton phases, and their connections to elasticity theory and gravity. We also discuss what is known about the dynamics and thermodynamics of fractons at non-zero density, before concluding with a brief survey of some open problems. 

\end{abstract}

\maketitle



\section{Introduction}
\label{intro}
The search for new states of matter is a central focus of condensed matter physics. Until recently, our understanding of the possible states of quantum matter was based largely on the traditional frameworks of broken symmetry \cite{landaulifshitz}, band theory, and Fermi liquid theory.  Today, frameworks are being developed that encompass topological phases of matter \cite{wenbook}, and new forms of quantum matter have been uncovered in non-equilibrium settings \cite{MBLARCMP, TC}. Recently, interest has been growing in a new class of quantum states of matter that does not wholly fit within any existing framework.
These are the fracton \cite{chamon,bravyi,castelnovo,haah,haah2,yoshida,fracton1,fracton2,sub,genem, KimHaah, williamson,hanCoupledLayer,sagarCoupledLayer,prem,mach,hsieh,slagle1,decipher,screening,nonabelian,balents,chiral,ppn,slagle2,albert,devakul,regnault,MaSchmitz, regnault2, albert2,leomichael, Gromov, SlagleChen, hanHiggs, bulmashHiggs, MaPretko} phases, and they constitute a new frontier for quantum condensed matter. 

The defining characteristic of fracton phases is that the elementary excitations thereof exhibit restricted mobility when acted upon by local operators -- they either cannot move without creating additional excitations (fractons), or else they can only move in certain directions (subdimensional particles). Meanwhile, {\it composites} of the elementary excitations are mobile particles, such that the elementary excitations themselves may be viewed as having `fractionalized mobility.' This restricted mobility endows fracton phases with glassy dynamics \cite{chamon, prem}, providing an intimate connection to ongoing research on glassy states of quantum matter. At the same time, gapped fracton phases in three spatial dimensions with periodic boundary conditions also exhibit a ground state degeneracy, that is {\it exponential} in linear system size, with distinct ground states being indistinguishable under local measurements \cite{chamon,bravyi,castelnovo,haah,haah2,yoshida,fracton1,fracton2}. The local indistinguishability of ground states is a property shared with topologically ordered systems, such as fractional quantum Hall liquids and gapped spin liquids. However, such more conventional systems have a constant ground state degeneracy in the thermodynamic limit.

Methodologically, two approaches have so far played a particularly important role in understanding fracton phases. One strategy is to construct and study exactly solvable spin models. This was the approach employed in Chamon's foundational paper \cite{chamon} and in the seminal early works \cite{bravyi,castelnovo,haah,haah2,yoshida,fracton1,fracton2}, and it makes liberal use of ideas and tools from quantum information theory. This approach is limited at present to systems in three spatial dimensions, and naturally gives rise to {\it gapped} fracton phases. A complementary approach pioneered by Pretko \cite{sub} studies instead gauge theories involving symmetric tensor gauge fields, and connects to distinct areas of physics, including elasticity theory \cite{leomichael} and gravity \cite{mach}. This approach naturally gives rise to {\it gapless} fracton phases, and also allows to extend the fracton phenomenology to two spatial dimensions. 

The fracton frontier thus sits at the confluence of multiple streams of research in theoretical physics. Perhaps in consequence, it has witnessed a remarkable flowering of research activity \cite{chamon,bravyi,castelnovo,haah,haah2,yoshida,fracton1,fracton2,sub,genem, KimHaah, williamson,hanCoupledLayer,sagarCoupledLayer,prem,mach,hsieh,slagle1,decipher,screening,nonabelian,balents,chiral,ppn,slagle2,albert,devakul,regnault,MaSchmitz,regnault2, albert2,leomichael, Gromov, SlagleChen, hanHiggs, bulmashHiggs}, with major new results appearing every few weeks. This review aims to provide a map of this frontier, or at least of those parts of it that we currently understand. It is structured as follows: in Section \ref{sec:gapped} we discuss {\it gapped} fracton phases. These are most naturally understood in terms of exactly solvable spin models, and we introduce two models of this type - the X-cube model and Haah's code. We highlight the key features of these models. We also explain how these models may be obtained from simpler models through layer constructions, dualities, or parton constructions, and how gapped fracton phases may be placed on general three dimensional manifolds. In Section \ref{sec:gaugetheories} we shift our attention to {\it gapless} fracton phases, which are most naturally described in the language of tensor gauge theories. We discuss the basic structure of these theories and also comment on how the gauge theory approach allows us to extend fracton phenomenology to two spatial dimensions, and how it connects to elasticity theory and gravity. In Section \ref{sec:fractonCMT} we move to discussing the dynamic and thermodynamic properties of fractonic matter at non-zero density. In Section \ref{sec:openquestions} we discuss some open questions that seem particularly important to us, and conclude.  We note that we have focused this review on {\it theoretical} work - the experimental study of fractons has barely begun, and we have not attempted to discuss it here. 

\section{Gapped fracton phases}
\label{sec:gapped}

\subsection{Solvable models for gapped fracton phases}
\label{sec:gapped-models}

We begin our exploration of gapped fracton phases by describing two exactly solvable models in three space dimensions that have played an important role in studies of fracton physics.  We first introduce the X-cube model of Vijay, Haah and Fu, which is perhaps the simplest fracton model \footnote{Historically, the X-cube model seems to have first appeared in Appendix D of Ref.\cite{CastelnovoXcube}}.  The X-cube model realizes an example of a ``Type I'' gapped fracton phase, where in addition to immobile fractons, there are also mobile sub-dimensional excitations.  We also discuss Haah's cubic code, which is in a ``Type II'' gapped fracton phase, where all the non-trivial excitations are immobile fractons.  Associated with this difference, the cubic code has a fractal structure that we will expose, while the X-cube model lacks fractal structure.  

\begin{figure}
(i) \includegraphics[width = 0.27\columnwidth]{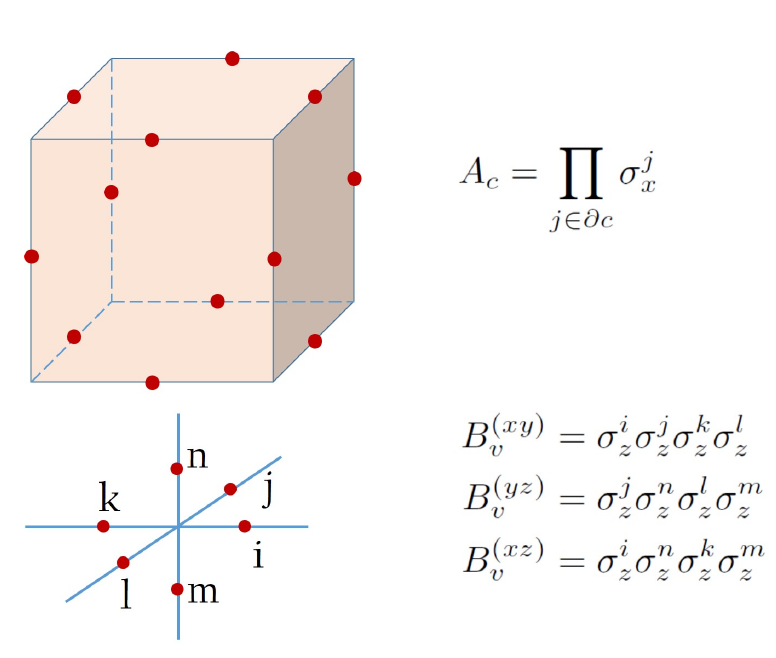}
(ii) \includegraphics[width=0.27\columnwidth]{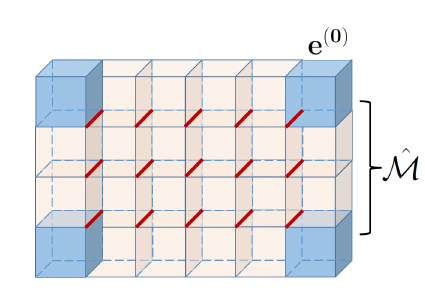}
(iii) \includegraphics[width=0.27\columnwidth]{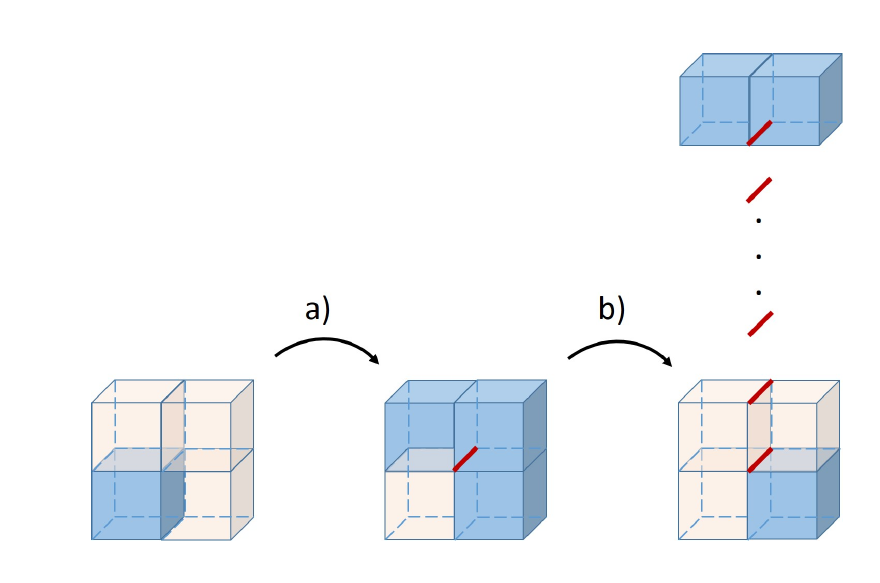}
\caption{Figures taken from Ref.\cite{prem} illustrating key features of the X-cube model. (i) An illustration of the stabilizer operators that make up the model, Eq.\ref{eq: Xcube}, using the notation $\sigma^\ell_z = Z_\ell$, $\sigma^\ell_x = X_\ell$. (ii) Acting with a rectangular membrane of $Z_{\ell}$ operators (red lines) creates a state where the four corners of the membrane (blue cubes) have eigenvalue $-1$ under the $A$ operator. These blue cubes are the fracton excitations of the model. (iii) A fracton can be moved by a local operator (red line), but only at the cost of creating additional excitations (two more `flipped' blue cubes). \label{fig:xcube}}
\end{figure}

In the X-cube model, spin-1/2 spins (qubits) are placed on the links of the
simple cubic lattice.  For each link $\ell$, we denote Pauli operators by $X_\ell$ and $Z_\ell$.  The Hamiltonian is 
\begin{equation}
H_{{\rm X-cube}} = - u \sum_c A_c - K \sum_{v, \mu} B^{\mu}_v    \text{,} \label{eq: Xcube}
\end{equation}
where $u, K > 0$, the first sum is over all elementary cubes $c$ and the second sum is over all vertices $v$ and directions $\mu = x,y,z$.  As illustrated in Fig.~\ref{fig:xcube},  $A_c$ is a product of $X_{\ell}$ over the 12 links in the boundary of the cube $c$, while $B^{\mu}_v$ is a product of $Z_{\ell}$ over the ``star'' of four links touching $v$ and lying in the plane normal to $\mu$.  All the terms in the Hamiltonian commute with one another, so the model is exactly solvable, and energy eigenstates can be labeled by the $\pm 1$ eigenvalues of the $A_c$ and $B^{\mu}_v$ operators.  The X-cube model is reminiscent of Kitaev's $\mathbb{Z}_2$ toric code, and like that model is a stabilizer code, which  means that every term in the Hamiltonian is a product of Pauli operators.

For an $L \times L \times L$ system with periodic boundary conditions, the ground state degeneracy ${\rm GSD}$ grows exponentially with $L$.  It was proved rigorously that for $L$ even, $\log_2 {\rm GSD} = 6 L - 3$.  The degenerate ground states cannot be distinguished by any local measurements, and the model is topologically ordered in that sense.

The X-cube model has two types of excitations.  The immobile fracton excitations reside on isolated cubes with $A_c = -1$.  Acting on a ground state with $Z_\ell$ creates four fractons on the cubes touching $\ell$, and four well-separated fractons can be created at the corners of a membrane operator that is a product of $Z_\ell$ over a rectangular region in e.g. the $\langle 1 0 0 \rangle$ plane.  In this sense, we can think of the X-cube model fractons as defects at the corners of rectangular membranes. In order to move a fracton, one must create two additional fractons at the other corners of a membrane; these gapped excitations cost energy, so the consequence is that a single isolated fracton is immobile even when generic local perturbations are added to the X-cube model (see Fig.\ref{fig:xcube}).

The other excitations of the X-cube model are one-dimensional particles that reside at vertices where \emph{e.g.} $B^{x}_v = B^y_v = -1$ and $B^z_v = 1$.  These excitations are created by acting with a string of $X_\ell$ operators, and can move, but only along a single axis parallel to the $x$, $y$ or $z$ direction; the example given moves along the $z$-direction.  The one-dimensional excitations are thus defects residing at the ends of rigid strings.  For a one-dimensional particle to turn a corner, it must emit an additional one-dimensional particle.

In general, we expect that any non-locally-creatable excitation of a bosonic system can be remotely detected by some process.  For instance, in $d=2$ topological order, an anyon $a$ can be detected via mutual statistics, where an anyon $b$ is moved in a circle around $a$.  Equivalently, because anyons are defects created at the ends of string operators, we can remotely detect $a$ by acting with an encircling loop of $b$-type string operator.  There is an analogous process in the X-cube model \cite{hanCoupledLayer}; an isolated fracton can be remotely detected by acting with a product of  string operators for one dimensional particles (\emph{i.e.} $X_\ell$) over the edges of a large rectangular prism.  The resulting statistical phase is $(-1)^{n_f}$, where $n_f$ is the number of fractons in the interior of the rectangular prism.  This example suggests that it should be possible to develop a theory of remote detection processes for fractons, generalizing more familiar ideas of braiding; this will be an exciting direction for future work.

\begin{figure}
\includegraphics[width=0.7 \columnwidth]{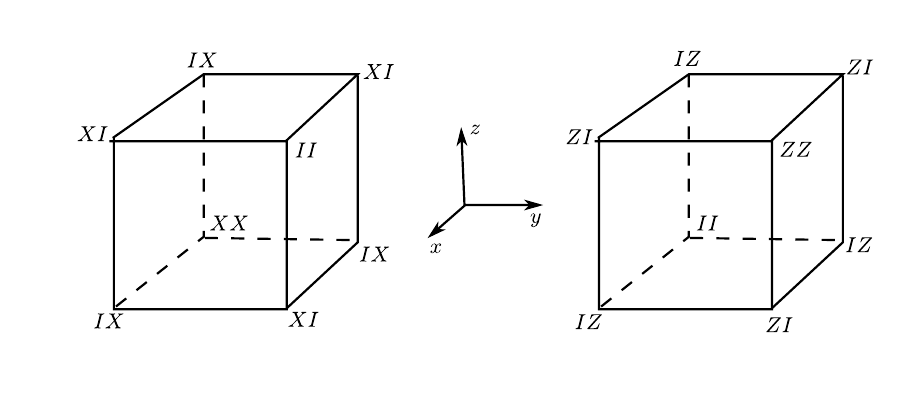}
\caption{Illustration of the two stabilizer operators that make up Haah's cubic code, Eq.~\ref{eq:Haah}. Figure taken from \cite{MaSchmitz}.}\label{fig:haah}
\end{figure}
Haah's cubic code is also defined on the simple cubic lattice, but now placing two qubits on each vertex.  We denote Pauli operators on each vertex by tensor products such as $X I = X \otimes I$ and $X Z = X \otimes Z$, where $I$ is the identity operator acting on one of the qubits.  The Hamiltonian is 
\begin{equation}
H_{{\rm Haah}} = -  \sum_c A_c -   \sum_c B_c \text{,} \label{eq:Haah}
\end{equation}
where both the $A_c$ and $B_c$ terms are defined on cubes $c$.  As shown in Fig.~\ref{fig:haah}, $A_c$ is a product of $Z$ Pauli operators, while $B_c$ is a product of $X$'s.  The ground state degeneracy of a $L \times L \times L$ system with periodic boundaries has a complicated dependence on the system size, but satisfies the simple upper bound $\log_2 {\rm GSD} \leq 4 L$ \cite{haah}.

\begin{figure}
\includegraphics[width=0.7 \columnwidth]{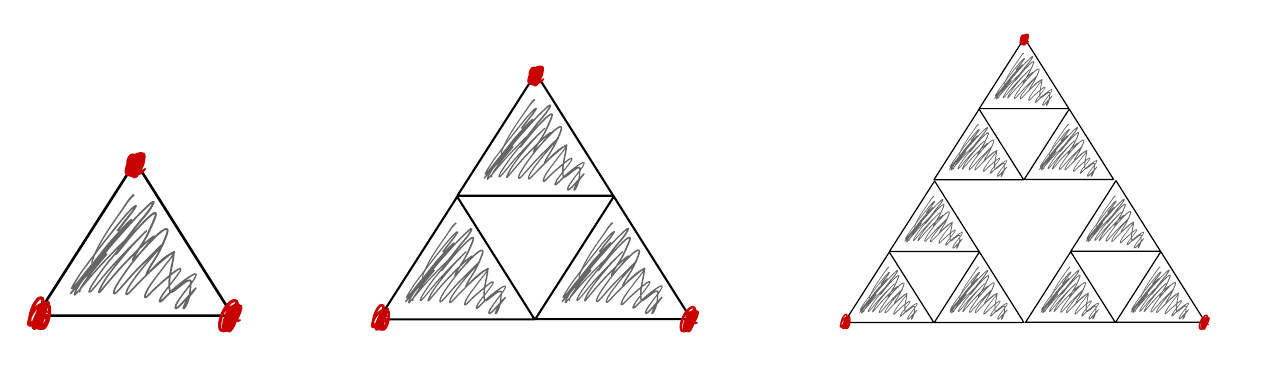}
\caption{Illustration of a two-dimensional analog of the  operators creating isolated fractons in Haah's code.  Three Ising-like excitations (red dots) are created at corners of a triangle (left).  To separate these excitations, triangles can be glued together, forming a fractal object (middle and right). A construction similar to this appears in Ref.\cite{castelnovo}}
\label{fig:triangle}
\end{figure}

Haah's code has a ``self-duality'' symmetry that exchanges the $A_c$ and $B_c$ terms, which is realized by a combination of  lattice inversion about a cube center, a swap of the two qubits on each lattice site, and unitary rotation that sends $X \to Z$ and $Z \to -X$ for each qubit.  Therefore it is enough to focus on the $A_c = -1$ excitations, which are immobile fractons.  Acting with $XI$ or $IX$ creates four excitations in two different tetrahedral patterns on neighboring cubes.  These four excitations can be separated far apart by ``gluing'' elementary tetrahedra in a fractal pattern, and the cubic code fractons are defects residing at the corners of a fractal object.  An analogous geometrical construction, based on triangles in two spatial dimensions, is illustrated in Fig.~\ref{fig:triangle}.

Conventional topological phases can be characterized by their topological entanglement entropy, which is a non-local contribution that can be extracted from the entanglement entropy, and related to universal properties of a topological phase.  This has been generalized to fracton models \cite{decipher, MaSchmitz, regnault2, albert2}, where a variety of non-local contributions to the entropy have been studied, that have been argued to be a robust property of the underlying gapped fracton phase.  A general feature of these ``topological'' entanglement entropies is the presence of a term proportional to the linear size of the subregion considered.  At least for the X-cube model, this term can be understood in a loop gas picture of the ground state.  In the $Z$ basis, the ground state can be viewed in terms of strings of links with $Z_\ell = -1$, which are constrained to form closed loops in every $\{1 0 0 \}$ plane (\emph{i.e.} all planes symmetry-equivalent to a plane normal to the $x$-axis).  Each plane cutting through a region $A$ thus contributes a constant non-local contribution to the entanglement entropy $S_A$, which results in the linear term.

\subsection{Topological and geometrical response of gapped fractons}
\label{sec:response}

In more conventional topologically ordered phases, an important role is played by the response of the system to changes in the global spatial topology, as measured by the ground state degeneracy.  That is, conventional topologically ordered phases have a unique ground state when placed on a sphere, but acquire a ground state degeneracy when placed on a torus and other manifolds of non-trivial topology.  The degenerate ground states cannot be distinguished by local measurements or correlation functions of local operators.  Instead, the operators with non-trivial matrix elements within the ground state subspace are string-like or membrane-like operators that wind around the space.

Gapped fracton phases do have degenerate, locally indistinguishable  ground states, and are topologically ordered in that sense.  However, the ground state degeneracy depends on and grows subextensively with the system size, so that ground state degeneracy is not simply a probe of the global topology.  What, then, does the ground state degeneracy probe?  Given the key role  discrete lattice geometry plays via the restricted mobility of excitations, it is natural to guess, and recent works indicate, that ground state degeneracy probes a combination of topological and geometrical properties of the space on which a gapped fracton phase is defined.

It is also understood how to define generalizations of the X-cube lattice model, both on more general lattices \cite{slagle3} and on general three-dimensional manifolds \cite{SlagleChen}.  In both these works, intersecting stacks of two-dimensional surfaces are embedded in the larger space, and qubits reside on the edges where two surfaces intersect.  Vertices, where three surfaces intersect, always look locally like vertices of the cubic lattice.  This allows one to define the X-cube Hamiltonian on the resulting lattice.  In the manifold context, such a structure is introduced by a foliation of the manifold, and the ground state degeneracy depends both on the global topology and on the foliation. In fact, even on manifolds of trivial topology but non-zero curvature, there can be robust ground state degeneracy \cite{slagle2}.  In the future, it will be interesting to explore questions of response to topological and geometrical structure beyond the X-cube model.

\subsection{Generalized gauge theories and dual spin models}
\label{sec:mappings}

While the study of gapped fracton phases began with exactly solvable spin models, recent work has moved beyond this paradigm, which is essential if fracton states are to be understood as quantum phases of matter.  Vijay, Haah and Fu \cite{fracton2} and independently Williamson \cite{williamson} showed that some fracton models are equivalent to a kind of generalized lattice gauge theory, and moreover are dual to spin models with subsystem symmetries.  Ref.~\cite{fracton2} started from subsystem-symmetric spin models as a means to discover new fracton models, including the X-cube model.  Here, we instead emphasize the generalized gauge theories and dual spin models as alternate descriptions of fracton physics, that we expect to be useful in future work.  Indeed, gauge theory and dual spin model representations play an important role in the theory of gapped spin liquids and other  more conventional topological phases. This subsection is somewhat technical in nature and may be skipped by the inexpert reader.

We illustrate these ideas with the X-cube model.  The equivalence between this model and a generalized gauge theory exactly parallels the equivalence of Kitaev's $\mathbb{Z}_2$ toric code with $\mathbb{Z}_2$ gauge theory \cite{kitaev2003fault}.  We introduce a gauge theory Hilbert space by placing qubits both on links $\ell$ and cube centers $c$ of the simple cubic lattice, denoting Pauli operators of the link variables by $\tilde{X}_\ell$, $\tilde{Z}_\ell$, and of the site variables by $\tau^x_c$, $\tau^z_c$.  The gauge theory Hilbert space is not a tensor product of site Hilbert spaces, but instead obeys the local constraints
\begin{equation}
\tau^x_c = \prod_{\ell \sim c} \tilde{X}_\ell \text{,}
\end{equation}
where the product is over the 12 edges bounding the cube $c$.  Gauge-invariant operators are those that commute with the local constraints, and $\tau^z_c$ and $\tilde{Z}_\ell$ are not gauge-invariant.  We can think of $\tilde{X}_\ell$ as the generalized electric field, $\tilde{Z}_\ell$ as the gauge potential, and $\tau^x_c$ as the gauge charge density. $\tau^z_c$ is the operator creating a gauge charge at $c$.

The gauge theory Hilbert space is in fact isomorphic to the spin-model Hilbert space of the X-cube model, with operators in the two Hilbert spaces identified as follows:
\begin{eqnarray}
\tilde{X}_\ell &=& X_\ell \\
\tau^z_{c_1} \tau^z_{c_2} \tau^z_{c_3} \tau^z_{c_4}  \tilde{Z}_\ell &=& Z_\ell \text{.}
\end{eqnarray}
In the second equation, $c_i$ are the four cubes that have the link $\ell$ as an edge.  These identities allow us to express the X-cube Hamiltonian in terms of gauge theory degrees of freedom.  Taking advantage of the local constraint and adding a new term with coefficient $J$, we have 
\begin{equation}
H_{{\rm gauge}} = - K \sum_{v, \mu} \tilde{B}^{\mu}_v - u \sum_c \tau^x_c  -J \sum_{\ell} \tau^z_{c_1} \tau^z_{c_2} \tau^z_{c_3} \tau^z_{c_4}  \tilde{Z}_\ell \text{,} \label{eqn:hgauge}
\end{equation}
where $\tilde{B}^{\mu}_v$ is given in terms of $\tilde{Z}_\ell$ by the same expression as $B^{\mu}_v$ expressed in terms of $Z_\ell$.  In this theory, gauge charges are fractons; indeed, they simply correspond to the $A_c = -1$ excitations of the X-cube model.  The $J$ term corresponds to a perturbation $-J \sum_\ell Z_\ell$ in the X-cube model, which in the gauge theory is manifestly a ``hopping'' term for the fractonic matter fields. In the same language, the one-dimensional $\tilde{B}^{\mu}_v = -1$ excitations are gapped magnetic particles.

This generalized gauge theory representation of solvable fracton spin models is closely related to the work of Hsieh and Hal\'{a}sz, who used parton techniques as a means to identify new fracton models \cite{hsieh}.  The partons are matter degrees of freedom coupled to a generalized gauge field, and, similarly to the Kitaev spin model on the honeycomb lattice \cite{kitaev2006anyons}, the models they construct are solved exactly upon representing them in terms of partons.

To obtain the dual spin model, we pass to the ``flux-free'' sector where $\tilde{B}^{\mu}_v = 1$;  we can achieve this by taking $K$ large and treating the first term in Eq.~(\ref{eqn:hgauge}) as a constraint.  The resulting constraint is solved by $\tilde{Z}_\ell = \mu_{c_1} \cdots \mu_{c_4}$, where  the $\mu_{c_i}$ are Ising variables residing on the cubes touching $\ell$. These variables can be absorbed into $\tau^z_c$, and we obtain the spin model
\begin{equation}
H_{{\rm spin}} =  - \sum_r \tau^x_r  -J \sum_{\square} \tau^z_{r_1} \tau^z_{r_2} \tau^z_{r_3} \tau^z_{r_4} \text{,}
\end{equation}
where the cube centers are the same as sites $r$ of the dual lattice, and the last term is a sum over dual plaquettes, with $r_i$ the four sites at the plaquette corners.    This model is invariant under a planar symmetry, where $\tau^z_r \to f(r) \tau^z_r$, with $f(r) = -1$ for $r$ contained within any $\{1 0 0 \}$ plane, and $f(r) = 1$ otherwise.  Reversing the steps in the duality, as in the treatment of \cite{fracton2, williamson}, amounts to gauging the subsystem symmetry.

\subsection{Mechanisms for gapped fracton phases}
\label{sec:mechanisms}

It is often useful to understand the degrees of freedom of one phase in terms of some other phase.  For example, many quantum spin liquids can be understood by quantum disordering either a magnetically ordered state or a superconductor, while suppressing the proliferation of certain topological defects, which become gapped excitations in the spin liquid.  This provides both a useful theoretical description, and a mechanism by which one phase can emerge from a proximate phase.  We note that the question of whether two phases are separated by a continuous quantum phase transition is a more detailed issue; the existence of such a mechanism only raises the possibility of a continuous transition.

It has been shown that some gapped fracton phases, in particular the phase of the X-cube model, can be obtained by coupling together two dimensional layers of $\mathbb{Z}_2$ toric codes \cite{hanCoupledLayer, sagarCoupledLayer}.  One starts from three perpendicular stacks of $d=2$ layers arranged into a cubic lattice.  In a strong-coupling limit, one obtains the X-cube Hamiltonian within degenerate perturbation theory.  This can be understood at intermediate coupling in terms of a particle-string (or, $p$-string) condensation mechanism, where the $p$-strings that condense are formed from the particle-like flux excitations of the toric code layers, arranged along the length of the string.  Single charge (vertex) excitations of the toric code layers are confined by the $p$-string condensate, but bound states of charges in two perpendicular layers are free to propagate, and become the one-dimensional particles of the X-cube model.  Moreover, in this picture, the fractons arise as the gapped defects at the ends of a broken $p$-string.  

These ideas were used to introduce a $\mathbb{Z}_n$ generalization of the X-cube model \cite{sagarCoupledLayer}, and a generalization based on stacking double semion models \cite{hanCoupledLayer}.  The latter example is distinguished from the ordinary X-cube model by the presence of non-trivial mutual statistics between certain pairs of one-dimensional particles.  A similar coupled-layer construction also underlies one family of models proposed to support non-Abelian fracton excitations \cite{nonabelian}.  Beyond construction of new models, describing fracton phases in terms of $p$-string condensation provides useful insights by relating these phases to more familiar $d=2$ topologically ordered phases.  It is an open and intriguing question whether all gapped fracton phases can be obtained from stacks of $d=2$ topological orders via some kind of condensation mechanism, or whether there are some gapped fracton phases for which this is impossible.

A different type of mechanism for fractons is provided by the coupled-chain constructions of Hal\'{a}sz, Hsieh and Balents \cite{balents}, who obtain fracton models by coupling together critical spin chains.  Remarkably, their construction requires only two-spin interactions. While the effective Hamiltonian for their fracton model arises only at a very high order of perturbation theory, with a corresponding small energy scale, these ideas may point the way to more realistic fracton models.

Some mechanisms relating fracton phases to one another have also been discovered.  In particular, the X-cube fracton phase can be obtained from  certain rank-two ${\rm U}(1)$ symmetric-tensor gauge theories (described in more detail in Sec.~\ref{sec:gaugetheories}), via a Higgs mechanism where charge-two matter fields condense \cite{hanHiggs, bulmashHiggs}.  This is analogous to the situation with ordinary vector gauge theory, where a ${\rm U}(1)$ gauge theory becomes a $\mathbb{Z}_2$ gauge theory upon condensing charge-two matter \cite{fradkin1979phase}.  In the fracton case, an interesting feature is that starting from a ${\rm U}(1)$ tensor gauge theory with fractons is not a sufficient condition for the charge-two Higgs phase to also have fracton excitations; for instance, if one starts with the simplest tensor gauge theory on the cubic lattice and condenses charge-two matter, one obtains a theory that has the same topological order as four copies of the $d=3$ toric code.  This model can then be transformed into the X-cube model by a selective condensation of certain flux loops.  An alternate route to the X-cube model  is to first condense certain point-like magnetic monopole excitations to obtain a different ${\rm U}(1)$ tensor gauge theory \cite{hanHiggs}, in which one subsequently condenses charge-two matter to obtain the X-cube model \cite{hanHiggs, bulmashHiggs}.  The effect of charge-two condensation has also been studied in a family of rank-two ${\rm U}(1)$ tensor gauge theories, with a range of resulting gapped phases including X-cube fracton order, conventional topological order, and trivial phases \cite{bulmashHiggs}.

\section{Gapless fracton phases}
\label{sec:gaugetheories}
We now turn our attention to {\it gapless} fracton phases, many of which can be described in the language of {\it tensor gauge theory}.  Early works studied spin models that realize tensor gauge theories as low-energy effective theories \cite{Xu1, Pankov, XuWuPlaquette,  Xu2}.  More recently, the presence  in these theories of fractons and other excitations with restricted mobility was discovered by Pretko \cite{sub}.
 We introduce the key features of tensor gauge theories in Sec.\ref{sec:gauge}, where we also explain the connection to the spin models previously discussed, and how these theories provide a powerful, complementary, and intuitive perspective on the fracton phenomenon.   In Sec.\ref{sec:electrodynamics}, we analyze gapless fracton phases of tensor gauge theories, and discuss how some  fracton phenomenology may extend to {\it two} spatial dimensions. In Sec.\ref{sec: connections} we discuss some of the surprising connections between gapless fracton phases and two seemingly very different areas of physics: elasticity theory and gravitation. We make use throughout of the Einstein summation convention, whereby repeated indices are implicitly summed over. 

\subsection{Symmetric tensor gauge theories}
\label{sec:gauge}

To set the stage, recall that in conventional gauge theories (like electromagnetism), the gauge variable is a vector object $A_i$, from which one can obtain a vector magnetic field $B_i$ and electric field $E_i$. In the Hamiltonian formalism, which is most convenient for our present purposes, the electric field $E_i$ is canonically conjugate to the gauge field $A_i$, and the requirement that states must be invariant under gauge transformations leads to a constraint on the electric field, which is simply Gauss's law. Gauss's law $\nabla \cdot \vec{E} = \rho$ in turn encodes a local conservation law.   We have
\begin{equation}
\int_V \rho \, dV = \int_V \nabla \cdot \vec{E} \, dV = \int_{\partial V} \vec{E} \cdot \vec{dA}.
\end{equation}
From this it follows that charge is locally conserved -- the charge in a volume $\rho$ is fixed by an integral on its boundary.  This implies that charge cannot be locally created or destroyed, so that the dynamics within a region restricts to a `superselection sector' with a definite value of the charge.
However, conventional electromagnetism does not support anything like the fracton phenomenon. Can we build more restrictive constraints and superselection rules into a more elaborate gauge theory? 

To make progress in this direction, we move away from gauge theories of vector objects to gauge theories of tensors. Rank two tensors are sufficient for our present purposes, and we will restrict our discussion to the rank-two case in this review. At first glance the claim that rank-two tensors offer something new might seem surprising, since it is known that two-form gauge theories (which are gauge theories of antisymmetric tensors) do not yield fractons or sub-dimensional particles in three spatial dimensions.\footnote{Two form discrete theories are dual to conventional one form discrete gauge theories, while two form continuous gauge theories are unstable to confinement in three or fewer spatial dimensions \cite{Orland, Pearson}. Two-form non-compact ${\rm U}(1)$ gauge theory in three dimensions is dual to the quantum XY model \cite{SavitDuality}.} We consider instead \emph{symmetric} rank-two tensors. 

The simplest symmetric tensor gauge theory is written in terms of a rank two symmetric tensor gauge field $A_{ij}$ with a generalized gauge transformation $A_{ij} \rightarrow A_{ij} + \partial_i \partial_j \phi$, where $A_{ij}$ is a real valued field and $\phi$ is a real valued scalar function.  In the Hamiltonian formulation, the conjugate variable to the gauge field is a rank two symmetric tensor electric field $E_{ij}$. The requirement of gauge invariance of states imposes on this conjugate variable a generalized Gauss's law constraint 
\begin{equation}
\partial_i \partial_j E_{ij} = \rho \text{.}
\end{equation}
This Gauss's law encodes {\it two} conservation laws. Not only is charge conserved (as in vector gauge theories), but so is {\it dipole moment}. The conservation of dipole moment follows simply from the Gauss's law, integration by parts, and the divergence theorem, since 
\begin{equation}
\int_V x_i \rho \, dV = \int_V x_i \partial_j \partial_k E_{jk} dV = \mathrm{surface \; terms \; on \;} \partial V \text{.}
\end{equation}
Thus the integrated dipole moment is fixed by an integral over the boundary of the volume, and is conserved in the sense that no local process can change the dipole moment. This conservation law in turn gives rise to a surprising local constraint: isolated charges {\it cannot move} since any movement of an isolated charge would change the dipole moment. Such a movement can only be accomplished by simultaneously creating an additional dipole, which costs energy and takes the system off energy shell. Therefore, the charges are fractons. Dipolar composites of fractons (e.g. a bound state of a positive and negative charge), meanwhile, are mobile objects. Thus, the generalized Gauss's laws associated with gauge theories of symmetric tensors naturally encode additional conservation laws, which impose superselection rules on the dynamics that naturally give rise to the fracton phenomenon. Tensor gauge theories in this manner provide a natural (and fully quantum mechanical) language for describing the fracton phenomenon.

We remark that, while our discussion is presented in terms of gauge theories in continuous space, one must bear in mind that these theories should be defined on a lattice.  Indeed, the lattice plays a much more significant role than in ordinary vector gauge theories.  For instance, on the lattice, the allowed dipole vectors of the scalar charge theory themselves form a lattice, whose points label different superselection sectors, so that the structure of the ``dipole lattice'' is a low-energy property of the deconfined phase (see Sec.~\ref{sec:electrodynamics}) where the dipoles exist as propagating excitations.  However, any definition of the scalar charge theory with continuous rotational symmetry would have a continuous infinity of allowed dipole vectors.  Therefore, if such a theory exists, it would not be correct to say it is regularized by the lattice scalar charge theory; rather, these are better viewed as two different theories.

It should be emphasized that tensor gauge theories are a framework -- there are any number of tensor gauge theories that could be written, with distinct gauge transformations and Gauss's laws, distinct Hamiltonians, and potential additional constraints such as tracelessness, and these can realize a multitude of different fracton phases. For a discussion of some additional examples, see \cite{sub}. Some of these variants also naturally realize subdimensional particles, which can move freely only in certain directions.  At least some gapped fracton phases, such as the X-cube model, can be described in terms of discrete tensor gauge theories on the lattice \cite{hanHiggs, bulmashHiggs}.  However, it is an open question how and whether all the generalized gauge theories of \cite{fracton2} are related to tensor gauge theories.

\subsection{Exploring gapless fracton phases}
\label{sec:electrodynamics}
The framework of tensor gauge theories allows us to explore gapless fracton phases.  Let us consider the scalar charge theory previously introduced, with a simple Maxwell Hamiltonian, 
\begin{equation}
H = \int d^3 r \big[ \mathrm{Tr}(E^{T} E) + \mathrm{Tr}(B^T B) \big] \text{,} 
\end{equation}
where the magnetic field is itself a tensor object defined by $B_{ij} = \epsilon_{iab} \partial_a A_{bj}$, with $\epsilon$ the Levi-Civita symbol. This theory has a gapless `photon' sector, and a gapped matter sector, the dynamics of which we have not written explicitly.

As for vector gauge theories, this theory can emerge from a lattice spin model, in the presence of a term that energetically enforces the Gauss law constraint at low energy.  In this setting, the theory is compact, meaning that the gauge potential is a $2\pi$-periodic variable, and there are also point-like gapped magnetic charge excitations.  Unlike the scalar electric charge, the magnetic charge transforms as a vector \cite{genem}.  An important point is that space-time magnetic instantons, which can lead to a gap for the photon in compact ${\rm U}(1)$ vector gauge theory in two dimensions, are not present here.  The stability of the gapless photon phase in this theory, and in many other tensor gauge theories, was demonstrated in \cite{Xu2}.

In \cite{sub} it was pointed out that charges in the gapless photon phase of the $d=3$ scalar charge theory display a phenomenon dubbed `electrostatic confinement.' A single charge produces electric fields (determined by the generalized Gauss's law), and the electrostatic energy stored in these fields diverges with system size. This means that isolated fractons are not finite-energy excitations, and instead are confined.  However, dipoles can be created with finite-energy cost, and interact with one another via a decaying power law potential. We emphasize that other tensor gauge theories with finite-energy fracton excitations have been identified; an example is the so-called traceless vector charge theory in three dimensions \cite{sub}.

The behavior of tensor gauge theories at finite temperature was discussed in  \cite{screening}. For the $d=3$ scalar charge theory, a crossover to a trivial phase was found, with a screening length that diverges as $\exp(m/T)$ at low temperatures, where $m$ is a mass scale for the theory.  Of course, even at zero temperature, since charge is a conserved quantity, we may prepare a state with as many charges as we want by introducing a suitable chemical potential, and by embedding the system in a suitable neutralizing `jellium' background (for more on this see Sec.\ref{sec:thermodynamics}).

The study of gapless fracton phases was extended in a new direction in \cite{chiral}, which added {\it topological} terms to the action for tensor gauge fields. The specific term added was a generalization of the famous $\theta \int \vec{E}\cdot{\vec{B}}$ term from conventional gauge theories. In conventional vector gauge theories in three spatial dimensions, such a term attaches electric charge to magnetic charges (Witten effect \cite{Witten}) and also gives rise to a Chern-Simons term on the boundary. What does such a term do to symmetric tensor gauge theories? 

To address the effect of topological terms, it is convenient to switch to the Lagrangian formulation. The action takes the form
\begin{equation}
S = \int_v dV \frac12 \left[(\dot{A}_{ij} - \partial_i \partial_j \phi)^2 - B_{ij} B_{ij} \right] + \frac{\theta}{4 \pi^2} \int_V (\dot{A}_{ij} - \partial_i \partial_j \phi) \epsilon_{iab} \partial_a A_{bj}  \text{,}
\end{equation}
where $\phi$ is a Lagrange multiplier. It may then be shown \cite{chiral} that a non-zero $\theta$ has the effect of attaching `fractonic' electric charges to the two ends of the magnetic charge vector, such that the magnetic charge vector also acquires an electric dipole moment, in a fractonic analog of the Witten effect. Such `charge attachment' also has the effect of transmuting the statistics of the mobile objects (e.g. dipoles), allowing us to work with either bosonic or fermionic dipoles. This observation will be important for the discussion in Sec.\ref{sec:thermodynamics}.

In a further illustration of the power of the tensor gauge theory approach, this formalism may also be used to obtain fracton phases in {\it two} spatial dimensions. In two or fewer spatial dimensions there appears to be some (not entirely understood) obstruction to constructing fractonic commuting projector Hamiltonians. However, there is no obstruction to writing down a symmetric tensor gauge theory. If one writes down a {\it compact} gauge theory then there is in principle a problem -- the theory undergoes confinement due to proliferation of instantons, as for compact vector ${\rm U}(1)$ gauge theory \cite{Polyakov}. However, this problem may be circumvented in one of two ways. If one works with a gapped fracton phase with a $\theta$ term (as discussed above), then on the two dimensional boundary of this three dimensional phase, the bulk $\theta$ term gives rise to a generalized `Chern-Simons' term which was argued to stabilize the boundary tensor gauge theory against confinement \cite{chiral}. Similar games may also be played in purely two dimensions, where it was argued that one can stabilize the theory against confinement by adding a Chern Simons term \cite{ppn}(thereby also rendering the theory chiral, and inaccessible via stabilizer codes).\footnote{However, \cite{Gromov} has argued that the apparent stability of the resulting theory may be an artifact of a linearization approximation.} More prosaically, however, if the goal is simply a stable fractonic phase in two spatial dimensions, one can just write down a non-compact tensor gauge theory. Such an approach was explored in \cite{leomichael}, which also uncovered an intimate connection to elasticity theory that we review in the next subsection. 

\subsection{Connections to elasticity theory and gravity}
\label{sec: connections} 

Gapless fracton phases are naturally described in terms of symmetric-tensor gauge theory. Symmetric tensors famously appear in two other contexts: in elasticity theory (stress and strain tensors), and in the theory of gravity (stress  and metric tensors). Could there be deep connections between these apparently disparate areas? The answer appears to be yes -- with exciting implications. 

  \begin{wrapfigure}{r}{0.5\textwidth}
\includegraphics[width = 0.5 \columnwidth]{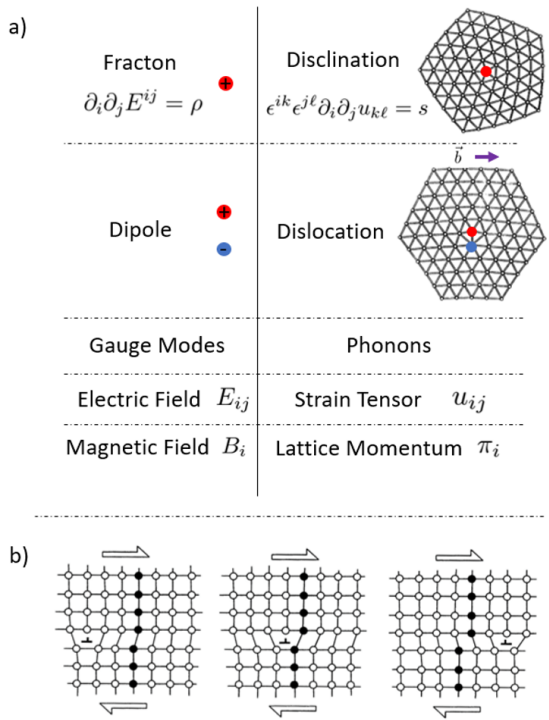}
\caption{\label{fig:elasticity} A summary of the duality mapping between fracton gauge theory and the quantum theory of elasticity in two spatial dimensions. Figure taken from Ref.\cite{leomichael}, with permission. }
\end{wrapfigure}
The connection between gapless fracton phases and the quantum theory of elasticity was worked out in \cite{leomichael}, which established a duality mapping between quantum elasticity theory and gapless fracton phases in two spatial dimensions. The gauge theory in question is a non-compact version of the scalar charge theory previously discussed, in two spatial dimensions. The duality maps fractons to disclinations of a quantum crystal, which share the property that they cannot move without creating additional excitations. Meanwhile, dislocations of the crystal  are mapped to dipoles in the fracton gauge theory, where the dipole and Burgers vectors are perpendicular.    Dipoles are free to move transverse to the dipole vector. Motion along the dipole vector locally changes the trace  $E_{i i}$.  Under duality, $E_{i i}$  corresponds to the density of vacancies and interstitials. In the absence of vacancies and interstitials, longitudinal dipole motion is forbidden, while in the presence of vacancies and interstitials, longitudinal dipole motion is suppressed and corresponds to dislocation climb (\emph{i.e.} motion normal to the Burgers vector) - see Fig.\ref{fig:elasticity}.  Finally, the gapless gauge modes of the fracton gauge theory are mapped to phonons in the language of elasticity, while the fractonic electric and magnetic fields map to the strain tensor and lattice momentum respectively. The duality mapping is summarized in Fig.\ref{fig:elasticity}.

The duality to elasticity theory provides powerful new insights into the physics of fracton phases. At the most basic level, this provides a mechanism for generating new stable phases of non-compact fractonic gauge theories, whose stability (as fully quantum theories) may be established by duality to elasticity.  Indeed Ref.\cite{leomichael} predicts several new fractonic phases in two spatial dimensions, via this duality. Whether the duality can be exploited in reverse, to use fracton physics to gain new insights into  the physics of quantum and classical crystals (or related more traditional areas of physics), and whether the duality can be extended away from two spatial dimensions, are exciting but still open questions. 

The tensor gauge theories introduced to describe fractonic models also exhibit phenomena reminiscent of gravity.\footnote{At first glance, the Weinberg-Witten theorem \cite{WeinbergWitten} may appear to rule out the possibility of gravity emerging from a tensor gauge theory ultimately regularized on a lattice. However, the Weinberg-Witten theorem assumes (a) Lorentz invariance and (b) the existence of asymptotic momentum eigenstates for particles. The typical setting for fractonic theories is not Lorentz invariant and, even more importantly, isolated fractons lack a momentum eigenstate description, so the theorem does not obviously apply. } This was recognized in \cite{mach}, where the following stimulating observations were made: First, it was observed that while isolated fractons are immobile objects, when multiple fractons are present in a system, they can move in a correlated fashion such that the additional conservation laws (\emph{e.g.} dipole conservation) are respected. Thus, while the ``inertial mass'' of an isolated fracton is infinite, the ``inertial mass'' of a fracton in a system containing other fractons is finite. Moreover, the closer fractons are together, the more local the operators needed to move them in a correlated fashion become, resulting in a lower effective mass. That is, the ``inertial mass'' of fractons depends on the density of other fractons in the same region of space, in analogy with the famous ``Mach principle'' in gravity \cite{mach1, mach2}.  Secondly, it was observed that fractons experience an effective {\it attraction}, since when close together, they can move about more freely. Finally, it was shown that the motion of fractons follows geodesics in the ``metric'' defined by the effective hopping matrix elements, such that the interaction between fractons appears to mimic gravity. 

Whether these analogs of gravitational phenomena are a sign of more precise connections between fracton physics and gravity remains to be seen.   For starters, fracton models with a gapped matter sector (such as the scalar charge theory), give rise to only a short range attraction between fractons. A long range attraction would require some method of stabilizing fracton theories where the matter sector is gapless. Furthermore, true Einstein gravity is non-linear, and this non-linearity is absent in the fracton theories that have currently been explored. Finally, the tensor gauge theories as written are not diffeomorphism invariant, as a theory of gravity must be. Despite these issues, it remains possible that insights from gravitation may illuminate the study of fractonic tensor gauge theories (or vice versa), and an exploration of these possibilities is an exciting avenue opened up by the study of gapless fracton phases. 

\section{Fractonic matter at non-zero density}
\label{sec:fractonCMT}
Thus far we have focused on fractons in isolation, interacting at most with a small number of other fractons. This may be viewed as the `particle physics of fractons' which, together with the work on fracton electrodynamics, constitutes the `standard model' of fractonic matter. A full understanding of fracton phases, however, requires also an understanding of the new emergent phenomena that may arise when fractonic matter is present at {\it non-zero density}. In this section we review what is currently known about the `condensed matter physics' of fractonic matter. In subsection \ref{sec:dynamics} we consider the behavior of fractonic spin models far from the ground state, and discuss the resulting {\it glassy dynamics}, and its connections to many body localization \cite{MBLARCMP}. In these models, fracton charge is only conserved modulo two, and as such fracton density and energy density cannot be decoupled. In subsection \ref{sec:thermodynamics} we therefore switch to the language of tensor gauge theories with conserved $U(1)$ charge. These theories allow for chemical potentials which couple both to fracton density and to the density of fractonic dipoles. By tuning these chemical potentials, one can turn on a finite density of fractons (or fractonic dipoles), without going to finite energy density, and can thus investigate the ground state physics of fractonic matter. This also allows us to study theories where both the matter and the gauge sectors are gapless. 

\subsection{Dynamics of fractonic matter}
\label{sec:dynamics}
Let us consider the quantum dynamics of fractonic stabilizer codes, such as the X-cube model or Haah code. We want to consider quantum dynamics starting from simple initial conditions, such as a definite configuration of excitations in real space.
 However, in stabilizer codes, any configuration of local excitations labeled by stabilizer eigenvalues is an eigenstate, and has trivial dynamics.  
 To make the question interesting, one must allow for arbitrary local perturbations of the stabilizer Hamiltonian, requiring however that the typical energy scale of these perturbations be much smaller than the `gap scale' $W$ in the stabilizer code, such that there is no phase transition. One can then study quantum Hamiltonian dynamics starting from an eigenstate of the unperturbed stabilizer code, with some definite real-space configuration of excitations. 
 
If the system is initialized in a sector with a vanishing density of excitations, then it follows from the superselection rules of the theory that the fractons must be immobile, since any movement of the fractons creates additional excitations and changes the energy of the state. Quantum Hamiltonian dynamics will then keep these fractons close to their initial positions, such that the system preserves forever a memory of its initial condition in local observables. As such, the system exhibits a form of `quantum localization,' at least in the subspace with vanishing density of excitations (colloquially, `zero temperature')  \cite{KimHaah}. Does this localization survive to non-zero energy densities? 

If the system is prepared at non-zero energy density then the behavior is considerably richer, since in this case there exists a `heat bath' internal to the system, made up of the mobile composite degrees of freedom, from which one may borrow energy to place fracton hopping back `on energy shell.' This problem was investigated in detail in \cite{prem}, and it was found that in the X-cube model, fractons can hop by `hitching a ride' on a two-fracton composite (which is mobile in two dimensions). However, such hopping is only possible when there is a mobile composite around, and the mobility of single fractons is thus proportional to the density of mobile composites. If the initial state is a Gibbs state at small but non-zero temperature $T$, then the mobility of individual fractons is $\sim \exp(-W/T)$, where $W$ is the energy gap for mobile composites i.e. fractons have exponentially low mobility at low temperatures. Additionally, if the fracton and mobile-composite sectors are initially prepared at different temperatures, then one may show by writing down and solving classical rate equations describing energy transfer between the sectors that the {\it approach} to equilibrium is logarithmically slow over exponentially long times i.e. the temperature difference between sectors decays as $1/\log(t)$ over a time window of order $\exp(+W/T)$. It thus follows, both from the suppressed mobility of fractons, and from the logarithmically slow relaxation, that the X-cube model at low but non-zero energy density exhibits {\it glassy} dynamics. 

The behavior is even richer for Haah's code, in which the energy cost of moving a fracton a distance $r$ grows as $W \log(r)$, at least up to the inter-fracton distance. Since the amount of energy that must be borrowed from the heat bath of composites grows with $r$, it follows that for a fracton to move a distance $r$, a time $\exp(W \log(r)/ T) = r^{W/T}$ must elapse. At low temperatures, this constitutes strongly {\it subdiffusive} relaxation $\langle r^2\rangle \sim t^\eta$ with $\eta = 2T/W \ll 1$. This {\it subdiffusion} continues at least up to the time that it takes fractons to `find each other.' This `relaxation timescale' may meanwhile be estimated as follows: if the initial state is a Gibbs state at a temperature $T$, then the density of fractons is exponentially small in $T$, and the typical interfracton spacing is $r \sim \exp(W/T)$. Substituting into the above formula, we conclude that the relaxation time for Haah's code should scale as $\exp(\frac{W}{T} \log \exp (W/T)) = \exp(\frac{W^2}{T^2})$. This {\it superexponential} dependence of relaxation time on temperature is reminiscent of `almost many body localized' systems \cite{GN, MirlinMuller}, and also of kinetically constrained models of classical glasses \cite{parabolic}. While these results were obtained in \cite{prem} using methods of closed system quantum dynamics, similar results also obtain if one considers a system in contact with an external heat bath, in which case the Lindblad formalism may be applied \cite{SivaYoshida}. 

The apparent similarity to classical glasses is far from accidental. Indeed, the dynamical rules governing the relaxation of fracton models are analogous to those of `kinetically constrained' theories in the `East model' class (for a review see e.g. \cite{Chelboun, Ritort} and references contained therein). These classical models also exhibit `dynamical facilitation,' whereby excitations (fractons) can move only in the vicinity of other excitations (mobile composites), and such `dynamical facilitation' provides a well known and robust route to glassy behavior (see e.g. \cite{Chandler, Biroli} and references contained therein). However, in kinetically constrained models these dynamical constraints are imposed `by hand,' whereas for fracton models they emerge naturally from Hamiltonian dynamics. Indeed, the earliest fracton models \cite{chamon, castelnovo} were specifically designed to have this property, generalizing classical `plaquette models' that also exhibit naturally emerging dynamical constraints (see e.g. \cite{Newmanmoore, GarrahanNewman}).  However, these early works also assumed contact with an external heat bath, whereas we now know \cite{prem} that fracton models also exhibit glassy dynamics under {\it closed} system quantum dynamics, providing an intimate connection to recent developments in quasi-many body localization in translation invariant systems ~\cite{MaksimovKagan, Schiulazbubbles, yaoglass, Papicstoudenmire, deroeck1, deroeck2, Garrahanglass}.

\subsection{Thermodynamics of fractonic matter}
\label{sec:thermodynamics}
In spin models such as the X-cube model and Haah's cubic code, fracton density and energy density cannot be independently controlled. In contrast, $U(1)$ tensor gauge theories allow for {\it independent} control of energy density and `fractonic charge density,' and thus hold out the promise of even richer behavior. The {\it dynamics} of fractonic matter, separating out the effects of non-zero charge density and non-zero energy density, have yet to be understood. However,  progress has been made \cite{ppn} in understanding the {\it thermodynamics} of fractonic matter at {\it zero} energy density but non-zero charge density, and we review this progress here. 

What is the ground state of a system containing a non-zero density of fractons \footnote{It is assumed that the system is embedded in a `fractonic jellium' background, so that there is a well defined thermodynamic limit. The theory is also implicitly defined on a lattice, restricting the allowed dipole moments.}? It turns out \cite{ppn} that this is determined by a delicate competition between kinetic and potential energy. On the one hand, fractons are able to move around when in close proximity to other fractons. For example, in a theory where fractonic behavior arises due to dipole moment conservation, two fractons could move in opposite directions, thereby leaving the total dipole moment unchanged. However, such motion requires an operator that acts on the scale of the inter-fracton separation. If we assume that the underlying Hamiltonian is purely local, then the amplitude of such `hopping' processes falls off exponentially with inter-fracton distance. This then generates a short range attraction between fractons (decaying exponentially with distance), in which fractons `like' to be close together so they can satisfy the kinetic terms in the Hamiltonian. However, there is also an electrostatic repulsion between fractons of like charge, which falls off as a power law with potential $V(r) = \alpha r^{-n}$, with $n$ determined by the precise theory (and Gauss law) under consideration. The competition between short range attraction and long range repulsion was addressed in \cite{ppn} (assuming weak $\alpha$) and it was shown that (i) for $n > 3$ the kinetic energy dominates, such that the fractons all cluster together in a single region of the system (ii) for $n < 0$ the repulsion dominates, and the ground state is a Wigner crystal of fractons, and (iii) for $0 < n < 3$ there arises an intermediate `micro-emulsion' phase in which fractons form clusters of finite size, and these clusters then form a Wigner crystal. 

Importantly, fracton phases exhibit additional conservation laws besides conservation of fractonic charge. For example, the `scalar charge theory' from Sec.\ref{sec:gaugetheories} exhibits a conserved {\it dipole moment}. In such theories, one may therefore `turn on' an additional (vector) chemical potential that couples to {\it fractonic composites} (elementary dipoles in the scalar charge theory), and thus address the phases of {fractonic composite matter} at finite density. Since fractonic composites are mobile particles even in isolation, one has now a richer variety of phases to explore. 

Ref.\cite{ppn} initiated the study of phases of fractonic composite matter by considering the simplest such question, in which one works with the scalar charge theory from Sec.\ref{sec:gaugetheories} and turns on a non-zero density of dipoles of a single orientation, while keeping the density of fractons zero. Problems with non-zero densities of fractons and dipoles, or non-zero densities of dipoles with multiple orientations, have yet to be addressed. Ref.\cite{ppn} also assumed for specificity that the dipoles were fermions (exchange statistics of dipoles can be transmuted as discussed in \cite{chiral}, and the equivalent problem with bosonic dipoles was not discussed). One then has a problem in which there are a finite density of mobile fermions (fractonic dipoles with a particular orientation), which interact via a $1/r$ repulsion in three spatial dimensions. The ground state, unsurprisingly, is conjectured to be a Fermi liquid, at least for not too strong repulsion. This seemingly prosaic ground state however supports an unexpected finite temperature transition. To demonstrate this, \cite{ppn} showed that while the `bare' inter-fracton potential is $\sim r$, in the presence of a Fermi surface of dipoles (of a particular orientation) it gets screened to $g \log(r)$. This logarithmic interaction then competes against entropy, such that above a critical temperature $T_c \sim g$ the dipoles `unbind' into fractons, destroying fracton order.

There is thus a rich structure and unexpected behavior to be found when one allows oneself to conside fracton systems at non-zero density, such that both the gauge and matter sectors are gapless. It should be emphasized moreover that most of this structure remains to be explored - Ref.\cite{ppn} only studied some of the simplest problems pertaining to thermodynamics of fractonic matter. There remain a plethora of open problems even in the thermodynamics while the dynamics has not even been touched on. These issues present a fertile field of exploration for future work. 

\section{Open questions and Conclusions}
\label{sec:openquestions}
The exploration of the fracton frontier is still in its early stages, and many important questions remain open (some of which we have touched on already). One obvious open question is: what types of fracton theories are possible, and how stable are they? For example, could the tensor gauge theory approach give access to fracton physics in one spatial dimension, as conjectured in \cite{ppn}? And if so, would the resulting theory be stable? Even in two and three spatial dimensions, are there qualitatively new types of fracton theories that have yet to be discovered? For example, a proposed non-Abelian extension of fracton theories was presented in \cite{nonabelian}. Another interesting open question is whether there exist {\it simple} lattice spin models that realize a fracton phase. The existing exactly solvable models are extremely complex, and an identification of simpler models realizing analogous physics is greatly to be desired. \footnote{Some progress in this direction was reported in \cite{slagle1, balents}.} A related question is whether all known fracton phases can be obtained from tensor gauge theories in some simple manner e.g. by condensing some set of objects. Another set of interesting questions seeks to exploit the connections of fracton phases to other areas of physics, such as quantum dynamics, quantum information theory, elasticity theory, and gravity. In particular, could the study of fractons provide new insights into these other, more traditional fields? Perhaps the most pressing open questions though, are: where could fractons be found in real materials, and how would we know a fracton phase where we see it? All these questions are wide open (as no doubt are others that we have overlooked) - there is much still to explore. 

In this review we have charted the current theoretical understanding of fracton phases of matter. It must be emphasized that despite the tremendous recent advances reviewed herein, the fracton frontier remains largely unexplored, and there remain no doubt new continents to discover. This is true both on the theoretical and, especially, on the experimental side, where the field is so young that any review seemed premature. We look forward to new researchers joining in the exploration of this new frontier and advancing our understanding of the field.
\newpage

{\bf Acknowledgements}  We are grateful for collaborations and discussions on fracton physics with Xie Chen, Andrey Gromov, Jeongwan Haah, Sheng-Jie Huang, Ethan Lake, Han Ma, Sid Parameswaran, Abhinav Prem, Michael Pretko, Leo Radzihovsky, Albert Schmitz and Hao Song. We also thank Leo Radzihovsky and Kevin Slagle for detailed feedback on the manuscript. This work was supported by the Foundational Questions Institute  (fqxi.org; grant no. FQXi-RFP-1617) through their fund at the Silicon Valley Community Foundation (RN), and by the U.S. Department of Energy, Office of Science, Basic Energy Sciences (BES) under Award number DE-SC0014415 (MH).

\bibliography{library}



\end{document}